\documentstyle[aps,twocolumn]{revtex}
\input epsf
\tightenlines
\begin{document}
\draft
\preprint{VECC-2000-}
\title{Scaling of single photon production
in hadronic collisions}

\author{Dinesh Kumar Srivastava}
\address{Variable Energy Cyclotron Centre, 1/AF Bidhan Nagar, Kolkata 
700 064, India}
\date{\today}
\maketitle

\begin{abstract}
Scaling of single photon production in $pp$ and $p\overline{p}$
collisions is studied. It is empirically observed the available
data scales $\sim\,\sqrt{s}/p_T^5$ for $x_T=2p_T/\sqrt{s}\,\leq\,0.1$ and
$\sim\,(\sqrt{s})^{3.3}/p_T^9$ for larger $x_T$. The NLO pQCD predictions
for $pp$ collisions at $\sqrt{s}$ of 200 and 5500 GeV, relevant for
 RHIC and LHC energies
are seen to closely follow this scaling behaviour. Implications
for single photon production in heavy ion collisions are discussed.

\end{abstract}
\narrowtext

Radiation of single photons in $pp$ and $p\overline{p}$ collisions
 have been studied for
a long time, for getting information about the partonic
distributions of nucleons and to test the applicability of
pQCD. A similar expectation is also associated with the study of
Drell-Yan process. In this connection the so-called Craigie fit~\cite{craigie}
to the Drell-Yan data, showing a scaling 
\begin{equation}
M^3\left(\frac{d^2\sigma}{dMdy}\right)_{y=0}=
 3 \times 10^{-32}\,e^{-15M/\sqrt{s}}
~ \mbox{\text{cm$^2$ GeV$^2$}}
\label{dy}
\end{equation}
has remained a very useful tool  for identifying the source of dileptons
in hadronic collisions.  Scaling relations are also useful in estimating
of the strength of `corrections' which cause a deviation  from the
expected behaviour.

Owens~\cite{owens} has discussed the possible scaling of the production
of single-photons in hadronic collisions. To leading order
in $\alpha_s$, single photons originate
from Compton ($qg\,\rightarrow\,q\gamma$) and annihilation 
($q\overline{q}\,\rightarrow\,g\gamma$)
processes, whose cross-sections, $d\sigma/dt$, have dimensions of 1/GeV$^4$.
This follows from the fact that the strong and electromagnetic coupling
constants are dimensionless and for massless partons no other (mass) scale
enters into the problem. This, Owens argued, can be used to
construct a scaling relationship for the invariant cross-section, 
$Ed^3\sigma/d^3p$, by combining the kinematic variable, $p_T$, $s$,
and $\theta$ (or the rapidity $y$), or equivalently, $p_T$, 
$x_T=2p_T/\sqrt{s}$, and $\theta$, etc., so that one could write
\begin{equation}
E\frac{d^\sigma}{d^3p}(A+B\,\rightarrow\,\gamma+X)=\frac{F(x_T,\theta)}{p_T^n}
\label{owens}
\end{equation}
with $n=4$ and $F(x_T,\theta)$  a dimensionless function.  This 
arguments needs to be refined to accommodate the fact that the strong
coupling constant $\alpha_s$ depends on the QCD scale parameter $\Lambda$
which has dimensions of momentum, and the structure functions depend
on the $Q^2$ at which they are sampled. This, along with
higher order terms,  would admit a more
complicated dependence on the momentum parameter $p_T$. It has been 
argued that such scaling violations, depending on the kinematic
region could raise $n$ to 6. 

The data for single photon production has been compiled and carefully
analyzed using NLO pQCD treatment by Vogelsang and 
Whalley\cite{vogel} and Aurenche et al~\cite{pat}.
 Our goal here is much more modest, we analyze them
empirically to look for scaling if any.
We see (Fig.~1) that, indeed, the data show different
scaling behaviour for the regions $x_T <0.1$ and $x_T>0.1$, as is evident from
the two lines drawn through them to guide the eye and to
 indicate the slope (i.e. the power of $p_T$) for a given $\sqrt{s}$. 

Next we perform a fit and find that to
a very good accuracy, the data show
a scaling, such that
\begin{equation}
\left(E\frac{d^3\sigma}{d^3p}\right)_{y=0}=6495\times\frac{\sqrt{s}}{p_T^5}
\mbox{\text{~~pb/GeV$^2$, $x_T<0.1$}}
\label{lowxt}
\end{equation}
which  varies as $F(x_T)/p_T^4$ and corresponds to $n=4$,
 suggesting that the scaling violations are
small in this kinematic region (see Fig.~2). Numbers varying by a few percent
are obtained in an unrestricted fit when the powers of $\sqrt{s}$ and
$p_T$ were used as free parameters. 

For the kinematic region $x_T > 0.1$ we find (see Fig.~3),
\begin{equation}
\left(E\frac{d^3\sigma}{d^3p}\right)_{y=0}=574.6\times\frac{(\sqrt{s})^{3.3}}
{p_T^{9.14}}
\mbox{\text{~~pb/GeV$^2$, $x_T>0.1$}}
\label{highxt}
\end{equation}
which varies as $F(x_T)/p_T^{5.8}$ and
corresponds to $n=5.8$ in the notation of Owens(Eq.\ref{owens}).  This 
indicates a large contribution of higher order processes and associated
deviation from the simple scaling at smaller $x_T$.

We digress a little to indicate that even though the E704 data~\cite{e704}
at $\sqrt{s}=$ 19.4 GeV have been included in the Fig.~3, they were not
included in the fitting procedure, which became unstable when this was done.
We also add that in the analysis of Vogelsang and Whalley~\cite{vogel}
only these data show a large deviation from the NLO pQCD results obtained
there.

 A comparison of the E704 data at $\sqrt{s}=$ 19.4 GeV with the
NA24\cite{na24} data at $\sqrt{s}=$ 23.75 GeV further shows that the former
has a result which is about 50\% larger at $p_T\,\approx$ 3.2 GeV 
which is very curious. The scaling seen here predicts that for a given
$p_T$, the production at 19.4 GeV should be a factor of two smaller,
 compared to its value at 23.75 GeV.

 Even though it is preposterous to argue about experimental data,
 it is tempting to note that the data at 19.4 GeV are a factor of
3.5 larger than `expected' on the basis of this scaling.  If it were indeed
 so, then the NLO pQCD results in Fig.~4 of Vogelsang and Whalley would
provide a perfect description to the `correct data', without any need of
inclusion of the so-called intrinsic $k_T$ effects\cite{cyw}. The inconsistency
of these data as well as the absence of requirement to include
intrinsic $k_T$ effects has been discussed in great detail by
Aurenche et al~\cite{pat}.

We have already noted that the NLO pQCD analysis by Vogelsang and Whalley
and Aurenche et al
has provided a reasonably accurate description of the data included in the
analysis here. Thus it would be fair to say that  the scaling behaviour
observed in the present work is a fair representation of the NLO pQCD
predictions for single photons from $pp$ collisions. 
(The slight difference between
the results for $pp$ and $p\overline{p}$ is neglected here. In any
case, the Compton term would dominate the contributions for not too
large values of $p_T$.)

In Fig.~4 we have plotted the NLO pQCD predictions of $pp$ scattering at
200 and 5500 GeV obtained in Ref.~\cite{jean}, which are relevant
for the experiments to be performed at RHIC and LHC. For the
higher energies the range of $p_T$ considered, limits $x_T$ to only 
smaller values and the scaling Eq.(\ref{lowxt}) provides a very good
description to the predictions. The range of $p_T$ considered at 200 GeV
is such that it spans both the low $x_T$ as well as the high $x_T$ regions
considered in the scalings seen here. It is gratifying to
note that the NLO pQCD  resultsi change over from the scaling 
Eq.(\ref{lowxt}) to that of Eq.(\ref{highxt}) as $x_T$ increases beyond 0.1.

What do these results mean for the recently measured single photon
data by the WA98 group for the $Pb+Pb$ collision at the CERN SPS?

We recall an interesting observation made some time ago by the
authors of Ref.~\cite{dks1}. Assuming that such heavy ion collisions
lead to the formation of quark-gluon plasma and assuming that the
system thus formed undergoes a 
boost-invariant expansion~\cite{bj}, 
 one can relate the particle rapidity density ($dN/dy$) to the
 initial time ($\tau_0$) and the temperature ($T_0$):
\begin{equation}
\frac{2\pi^4}{45\zeta(3)}\,\frac{1}{A_T}\frac{dN}{dy}=4 a
T_0^3\tau_0
\label{T0}
\end{equation}
where $A_T$ is the transverse dimension of the system and $a$ is decided
by the number of degrees in the plasma. 

It was pointed out ~\cite{dks1}, that the quantity
\begin{equation}
\frac{1}{A_T}\frac{dN}{dy} \approx 5~~ {\mbox{\text{fm}$^{-2}$}}
\label{dndy}
\end{equation}
for both $S+Au$  and the $Pb+Pb$ systems in the WA80~\cite{wa80}
 and the WA98~\cite{wa98}
experiments at the CERN SPS, we are offered a unique opportunity of
comparing two systems of different volumes which may have identical 
initial conditions! It is seen that if the transverse expansion
of the system does not play a significant role then for a given
 $\tau_0$, the only
scale in the system is provided by the temperature, for a baryon free
plasma.

If this reasoning is correct then the radiation of single photons
per unit transverse area would be identical. This leads to a
simple geometrical factor of $\sim$ 3.5 by which the data for $S+Au$
system can be scaled to get the results for $Pb+Pb$ system.

What about the contribution of prompt photons for the two cases?
The scaling behaviour of the
prompt photons seen here suggests that we may obtain the prompt
photon yield for the WA98 experiment as:
\begin{eqnarray}
\left(\frac{dN_{\text{prompt}}}{d^2p_Tdy}\right)_{PbPb}&=&
\left(\frac{17.4}{20}\right)^{3.3}\times \left(\frac{T_{PbPb}(b=0)}
{T_{SAu}(b=0)}\right)\nonumber\\
& &\times \left(\frac{dN_{\text {prompt}}}{d^2p_Tdy}\right)_{SAu}
\end{eqnarray}
where $\sqrt{s}_{NN}=$ 17.4 and 20 GeV for the WA98 and the WA80
experiments. This suggests that the prompt photon production for the
WA98 experiment can be obtained by multiplying the corresponding
contribution for the WA80 experiment by a numerical factor of $\sim$
3.43. This is quite close to the factor of 3.5 obtained earlier
for the thermal photon yield! 
In view of the above it is felt that the sum of thermal and prompt
photon productions for the two experiments should differ by a factor of
about 3.5!

In Fig.~5 we show the upper limit of the $S+Au$ multiplied
by this factor against the (upper limit and) the data for the $Pb+Pb$
system reported by the WA98 experiment. It is
a pity that the weekness of the this signal which is buried into
the huge back-ground of decay photons has resulted in the identification of
only the upper-limit of the single-photon production for the $S+Au$
system. Still, it is interesting to note that the upper limit measured
in the WA80 experiment is consistent with the excess of single photon
production obtained by the WA98 experiment. 

We have also shown a recent explanation of the data
data~\cite{dks2} using (corrected) two loop rates for photon production
from the QGP along with the contribution of hadronic reactions,
as well the prompt photons estimated by Wong and Wang~\cite{cyw} within
a pQCD with inclusion of effects of the intrinsic $p_T$ of the partons.

In brief, we have seen that the data for single photon production 
in nucleon-nucleon collisions can be broadly divided into two regions
$x_T\,<$ 0.1 and $x_T\,>$ 0.1. A scaling behaviour $\sim\,F(x_T)/p_T^4$ is
seen for low $x_T$ data, as expected from leading order pQCD, while
for larger $x_T$ a behaviour $\sim\,F(x_T)/p_T^{5.8}$ is observed, which 
indicates large corrections to the lowest order QCD results. It is
hoped that these observations may provide a useful guide-line for
identification of source of single photons as well as the extent of
corrections over the lowest order pQCD for these processes.

\section*{Acknowledgements} We thank Terry Awes, Charles Gale,
Juergen Schukraft, and Bikash Sinha for useful comments.

\newpage

\begin{figure}
\epsfxsize=3.25in
\epsfbox{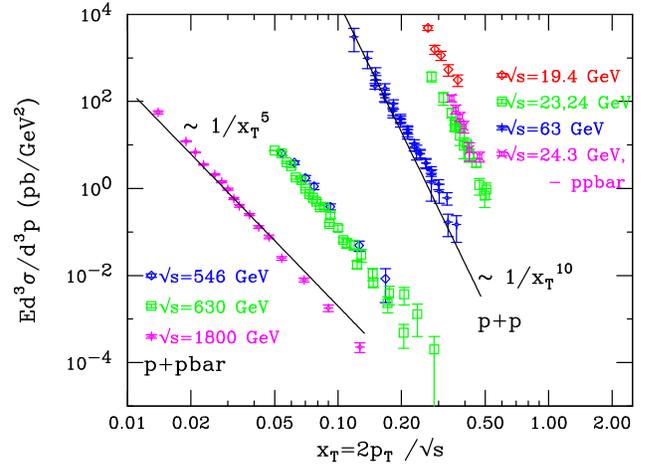}
\vskip 0.2in
\caption{
The available single photon data for \protect $pp$ and 
\protect $p\overline{p}$ taken from the compilation of Vogelsang 
and Whalley~\protect\cite{vogel}.  The large \protect $x_T$ data 
are taken from E704 experiment (19.4 GeV) \protect\cite{e704},
WA70 experiment (22.96 GeV) \protect\cite{wa70}, NA24 experiment (23.75 GeV)
\protect\cite{na24}, UA6 experiment (24.30 GeV) \protect\cite{ua6},
R110, R806, and R807 experiments (63 GeV) \protect\cite{r110,r806,r807}
and the UA6 experiment at 24.30 GeV for \protect $p\overline{p}$.
The small \protect $x_T$  data are limited to $p\overline{p}$ and are
taken from the UA1 and UA2 experiments (540 and 
630 GeV)~\protect\cite{ua1,ua2} and CDF and D0 experiments at 1800 GeV
\protect\cite{cdf,d0}.
}
\end{figure}

\begin{figure}
\epsfxsize=3.25in
\epsfbox{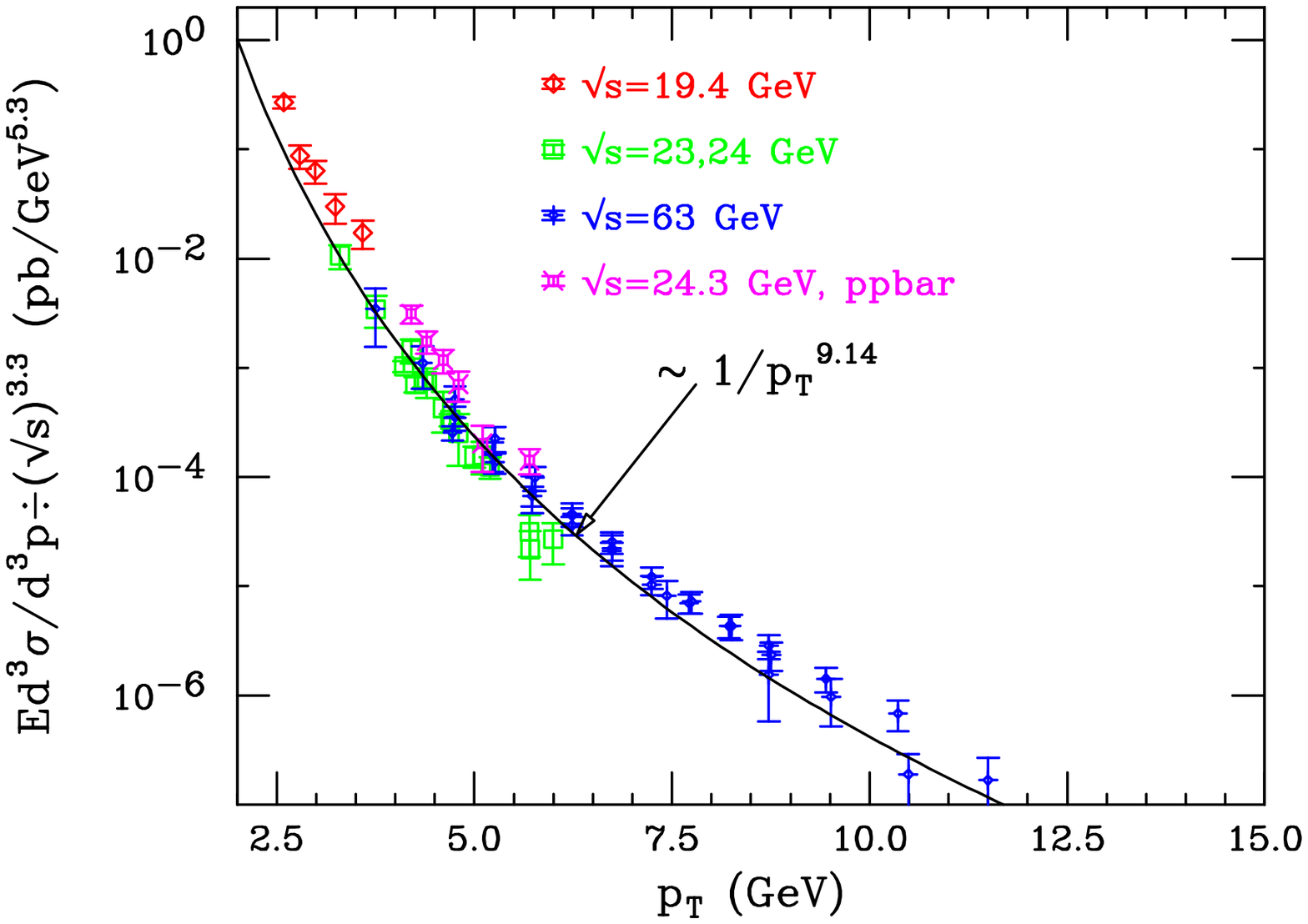}
\vskip 0.2in
\caption{ Fit to single photon data at large \protect $x_T$ using the scaling
 (Eq.\protect\ref{highxt}) obtained in the present work.
}
\end{figure}
\begin{figure}
\epsfxsize=3.25in
\epsfbox{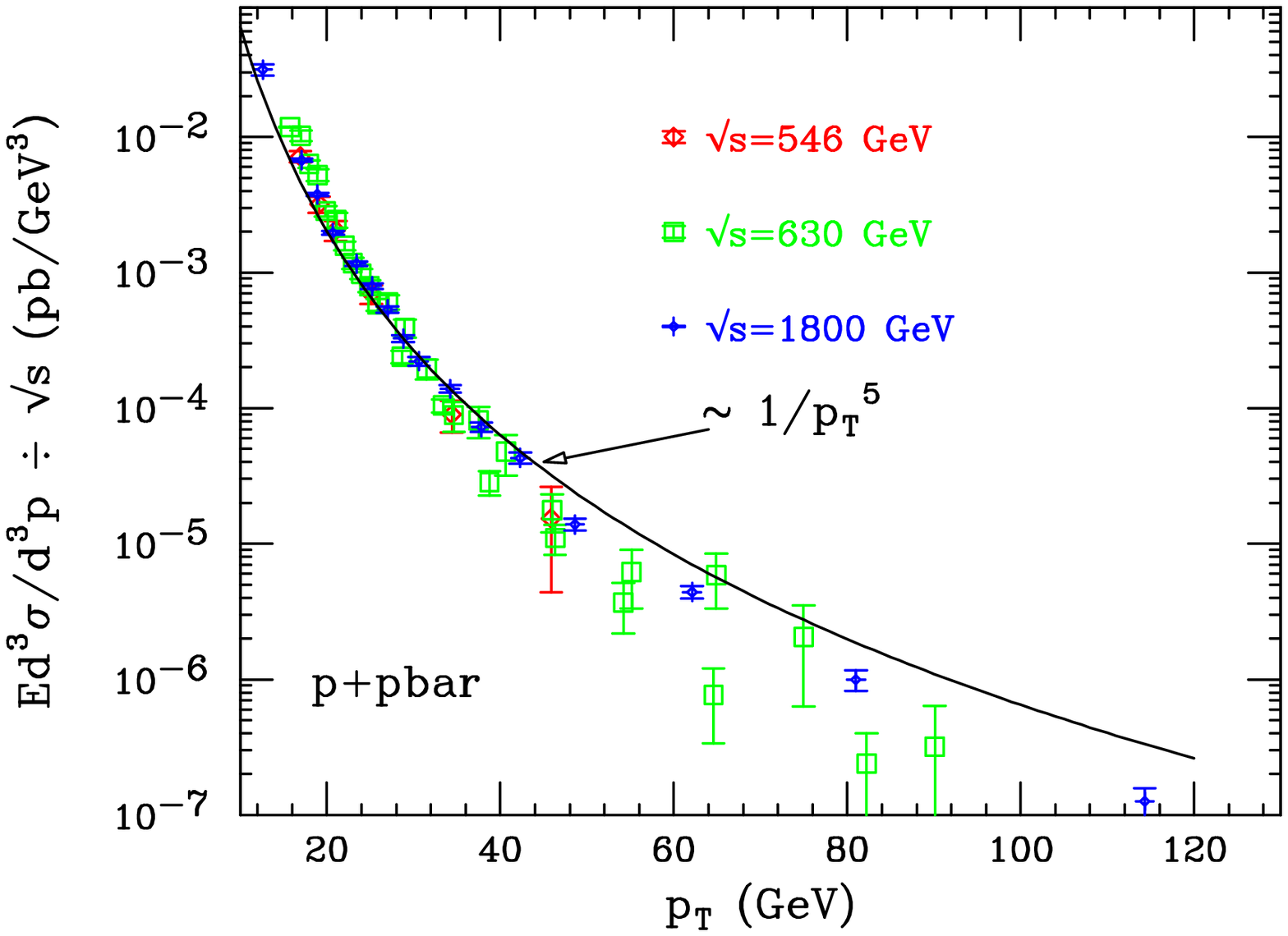}
\vskip 0.2in
\caption{
Fit to single photon data at low \protect $x_T$ using the scaling
 (Eq.\protect\ref{lowxt}) obtained in the present work.
}
\end{figure}
\begin{figure}
\epsfxsize=3.25in
\epsfbox{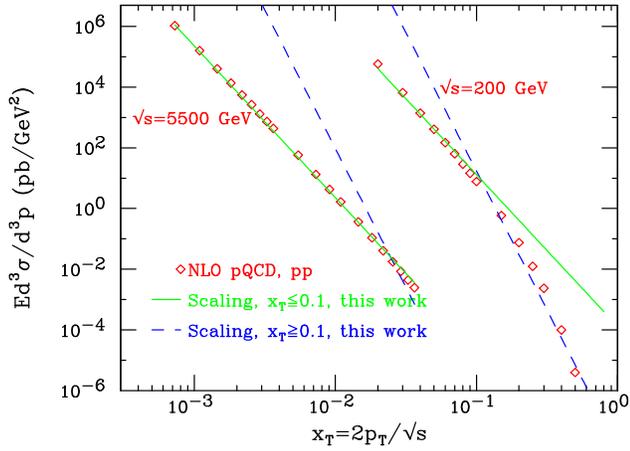}
\vskip 0.2in
\caption{
Test of the scaling observed in the present work against NLO pQCD
predictions\protect\cite{jean} for single photons at RHIC and LHC energies in 
\protect $pp$ collisions. 
}
\end{figure}
\newpage
\begin{figure}
\epsfxsize=3.25in
\epsfbox{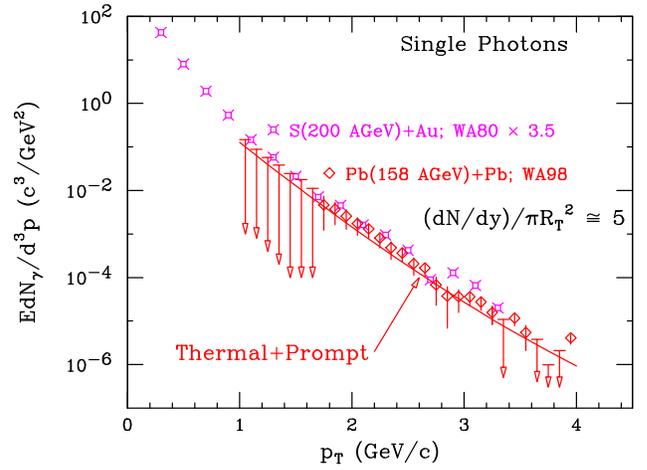}
\vskip 0.2in
\caption{
Single photon production observed in \protect $S+Au$ collisions (only
upper limits) in WA80 experiment~\protect\cite{wa80} and in \protect $Pb+Pb$
collisions in WA98 experiment~\protect\cite{wa98}. The WA80 `data' have been
rescaled using the scaling relation suggested in Ref.\protect\cite{dks1} and
implied by the relations (5) and (6) given in the text. The solid
curve gives the predictions of Ref.\protect\cite{dks2} suggesting 
a thermal source for these photons, while the dashed
curve gives the predictions based on the scaling observed in this work
for prompt photons.
}
\end{figure}

\end{document}